\documentclass[a4paper,11pt]{article}
 
\pdfoutput=1 

\usepackage{jinstpub} 
\renewcommand{\arraystretch}{6.5}
\title{\boldmath An Automatic Data Cleaning Procedure for Electron Cyclotron Emission Imaging on EAST Tokamak Using Machine Learning Algorithm}

\author[a]{C. Li,}
\author[a,1]{T. Lan,}
\author[a]{Y. Wang,}
\author[a]{J. Liu,}
\author[b]{J. Xie,}
\author[b]{T. Lan,}
\author[b]{H. Li}
\author[a,c]{and H. Qin \note{Corresponding author.}}

\affiliation[a]{Department of Engineering and Applied Physics, School of Physical Sciences, \\University of Science and Technology of China, No. 96 Jinzhai Road, Hefei, China}
\affiliation[b]{School of Physical Sciences, University of Science and Technology of China, \\No. 96 Jinzhai Road, Hefei, China}
\affiliation[c]{Plasma Physics Laboratory, Princeton University,
\\Princeton, NJ, 08543 U.S.A.}
\emailAdd{lanting@mail.ustc.edu.cn}

\abstract{A new data cleaning procedure for electron cyclotron emission imaging (ECEI) of EAST tokamak is developed. Machine learning techniques, including Support Vector Machine (SVM) and decision tree, are applied to identifying saturated, zero, and weak signals of ECEI raw data, which not only reduces the effort of researchers for data analysis, but also improves the accuracy of data preprocessing. Proper training sets are sampled using massive raw ECEI data from the EAST tokamak. Optimal window size of temporal signal, kernel function, and other model parameters are obtained by model training. With the optimized parameters, the recognition rates of saturated, zero, and weak signals in raw data are 99.4\%, 99.86\%, and 99.9\%, respectively, which proves the accuracy of this procedure.}


\proceeding{N$^{\text{th}}$ Workshop on X\\
  when\\
  where}

\begin{document}
\maketitle
\flushbottom
\section{Introduction}
\label{sec:intro}

Electron Cyclotron Emission Imaging (ECEI) has been introduced as a useful diagnostic method for detecting the two dimensional (vertical and horizontal) distribution of electron temperature ($T_e$) in different tokamaks~\cite{a,b}. Based on ECEI data, formation of magnetic islands and MHD instabilities can be investigated~\cite{c,4}.
ECEI data thus play a critical role in the study of magnetic confined plasmas. However, due to the complexity of measurement environment and the limitation of acquisition range, raw ECEI data contain lots of invalid signals, including saturated signals, zero signals, and weak signals. The appearance of these invalid signals increases the complexity of data analysis. Data cleaning is inevitable before further analysis. Traditional approaches of classifying invalid ECEI data mainly rely on human eyes, which is obviously slow and subjective to human errors.  On the EAST tokamak, one discharge generates 7.6 GB ECEI data, and massive ECEI data have been accumulated. An automatic data cleaning tool is needed.

As a powerful technique for data analysis, machine learning has entered the field of plasma physics.  For example, the position of the magnetic probe and inversion radius can be determined based on neural network algorithm~\cite{5,6}. Researchers are also seeking ways to predict the disruptions of plasmas in large fusion devices via machine learning method~\cite{7}. Compared with traditional methods, the results of machine learning technique improves with the increase of training data. At the same time, the human labor can be replaced by computers, which accelerates the process of discovery.

In this paper, a new data cleaning procedure using machine learning algorithms, is developed for ECEI system of the EAST tokamak. The system is a muti-channel system, and the similarity of one shot data is not obvious. Therefore, several machine learning methods are combined to identify different kinds of signals rather than using preference-based performance measures~\cite{8}. Our procedure starts from the window separation which divides an ECEI signal into segments with unified length. Then, Support Vector Machine (SVM) and decision tree are used to analyze these segments and classify the properties of raw data.  SVM has a unique advantage in high-dimensional pattern recognition~\cite{9}, and decision tree is easy to use and powerful for addressing optimization problems~\cite{10}. We use SVM to identify saturated signals, and apply the method of decision tree to classifying zero signals and weak signals. In order to validate models, a five-fold cross validation is used, which can efficiently ease the problem of over-fitting. To enhance the reliability of the procedure, proper training sets are sampled from massive raw ECEI data of the EAST tokamak. System parameters including the window size and the kernel function are adjusted to optimize the model. It is found that the performance of model is sensitive to the size of window. To be specific, smaller window size implies larger computational complexity, and larger window size reduces the effectiveness of features. It is also found that the polynomial kernel function can be effectively trained by SVM.   With these techniques, accuracies of identifying saturated signals, zero signals and weak signals reach 99.4\%, 99.86\%, and 99.9\% respectively. The training time will be limited to several seconds if GPU is used. 

This paper is organized as follows. Section~\ref{sec:2} briefly introduces the physical background of the ECEI data. In Section~\ref{sec:3}, details of the classification procedure are provided. Sample sets are shown in Section~\ref{sec:4}. Section~\ref{sec:5} shows results of the validation. And a summary is given in Section~\ref{sec:6}.

\section{The ECEI data}
\label{sec:2}
Each set of the ECEI data in the EAST tokamak consists of 384 channels (24 vertical and16 horizontal). Each channel detects the electron cyclotron radiation in the tokamak, and the radiation is mixed with the local oscillation (LO) frequency on the antenna array and down-converted to the IF~\cite{11,12}. After the signal is received by the intermediate frequency system, it is amplified with band-pass filters and converted to an analog signal. The final radiation image is obtained through the data acquisition card.

The signal lasts for ten seconds and ranges from -1 V to 1 V. The range is 0 V to 2 V if zero drift is processed. Typically, the range of saturated signals is from -1 V to 1 V, and the range of zero signals and weak signals is from 0 V to 2 V. By analyzing the electron temperature fluctuation ($\sigma T_e$), sawtooth instability can be studied~\cite{13,14,15}. In general, there are three types of invalid data: saturated signals, zero signals, and weak signals. According to previous experience, $T_e$ profiles of saturated signals exceed the range, insufficient attenuation generates saturated signals, and zero signals can be considered that they are almost all noise. The following facts are also the characteristics of the invalid data: $T_e$ profiles of zero signals are very close to 0 V; zero signals are due to the error of the antenna route; the signal-to-noise ratio of weak signal is stronger than that of zero signal but weaker than that of normal signal. 

\begin{figure}[htbp]
\centering 
\includegraphics[width=.8\textwidth]{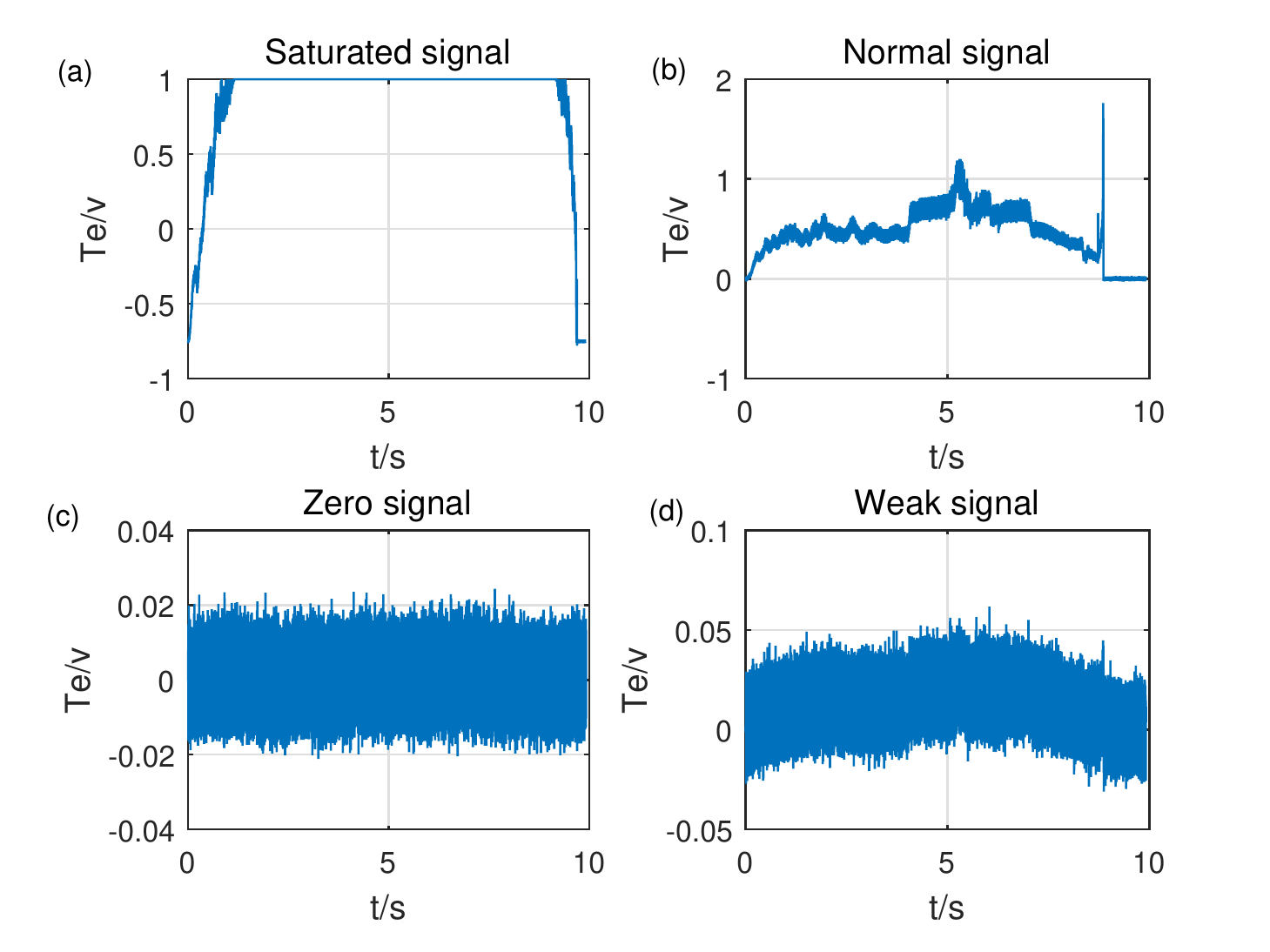}
\caption{\label{fig:1}Typical patterns for four kinds of signals in the ECEI measurement: (a) the pattern of a
saturated signal, (b) the pattern of a normal signal, (c) the pattern of a zero signal, and (d) the pattern
of a weak signal.}
\end{figure}

Figure~\ref{fig:1} shows typical patterns of saturated signals, normal signals, zero signals, and weak signals. The signal in figure~\ref{fig:1}(a) is considered as a saturated signal because it reaches the range during 4s and 6s. The signal in figure~\ref{fig:1}(c) is a zero signal because its baseline is essentially at 0 V. The amplitude of the signal in figure~\ref{fig:1}(d) differs slightly from that of the noise. So it is regarded as a weak signal caused by the poor signal-to-noise ratio.

\section{Procedure for the classification of signals}
\label{sec:3}
\begin{figure}[htbp]
\centering 
\includegraphics[width=.8\textwidth]{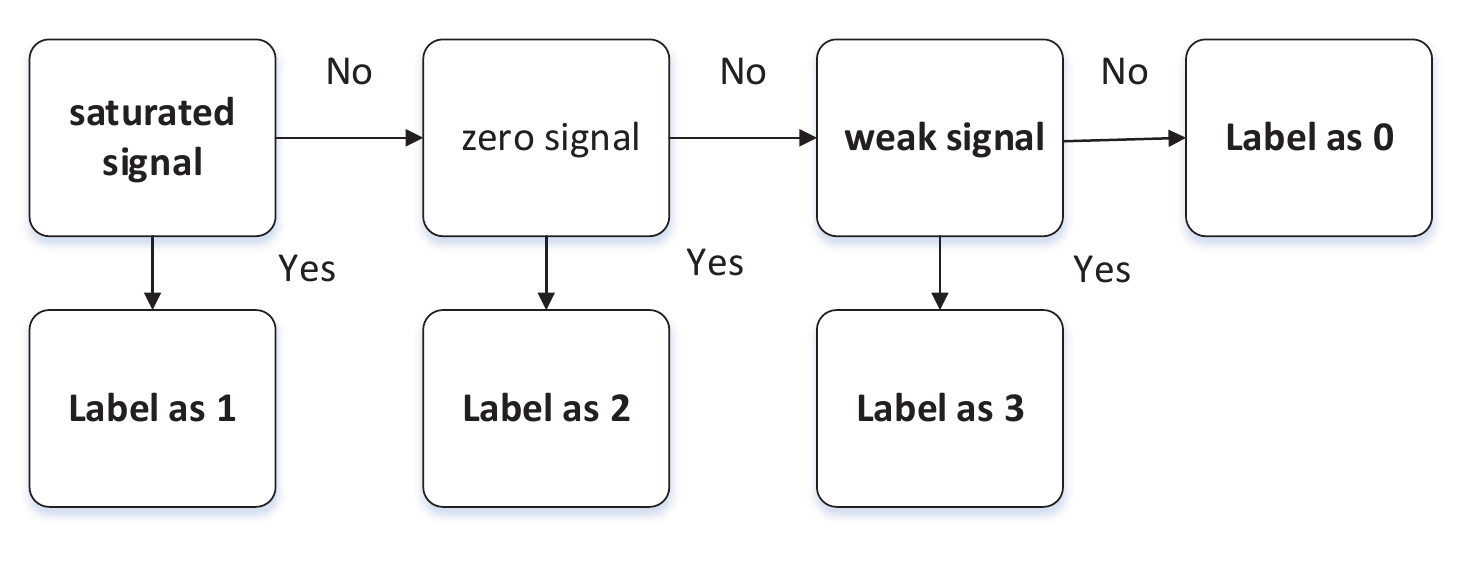}
\caption{\label{fig:2}The flow chart of the classifier.}
\end{figure}

To classify four kinds of signals above, a classifier is set up. The flow chart of the classifier is shown in figure~\ref{fig:2}. Saturated signals, zero signals, and weak signals are identified in sequence. Saturated signals, zero signals, weak signals, and normal signals are marked as 1, 2, 3 and 0, respectively.

\subsection{Identification of saturated signals}
The feature of saturated signals is that there will be a period of saturation. The length of duration as a parameter is optional. We select the duration as 0.6 seconds. The model of identifying saturated signals will be trained by SVM. The distance between each signal and 1 V is defined as the parameter of the sample. Sample sets are divided into two categories. One is saturated, and the other is unsaturated. Each sample is labeled, 1 means saturated, and 0 means unsaturated.

\subsubsection{Pretreating}
Before training the model, parameters of signals are obtained according to pretreating. The signal is divided into fifty windows. For each window,  a distance L relative to the range is defined by the standard deviation 
\begin{equation}
\label{eq:1}
L = \sqrt{(\sum_{i=1}^{n}(l_{i}-1)^{2})/n},
\end{equation}
which measures the fluctuation relative to the mean. If we set the expectation of the formula to 1v, it will represent fluctuations relative to the range.  In \eqref{eq:1}, $l_{i}$ is the electron temperature at the i-th time step and n is the total number of time steps in each window. 

\begin{figure}[htbp]
\centering 
\includegraphics[width=.8\textwidth]{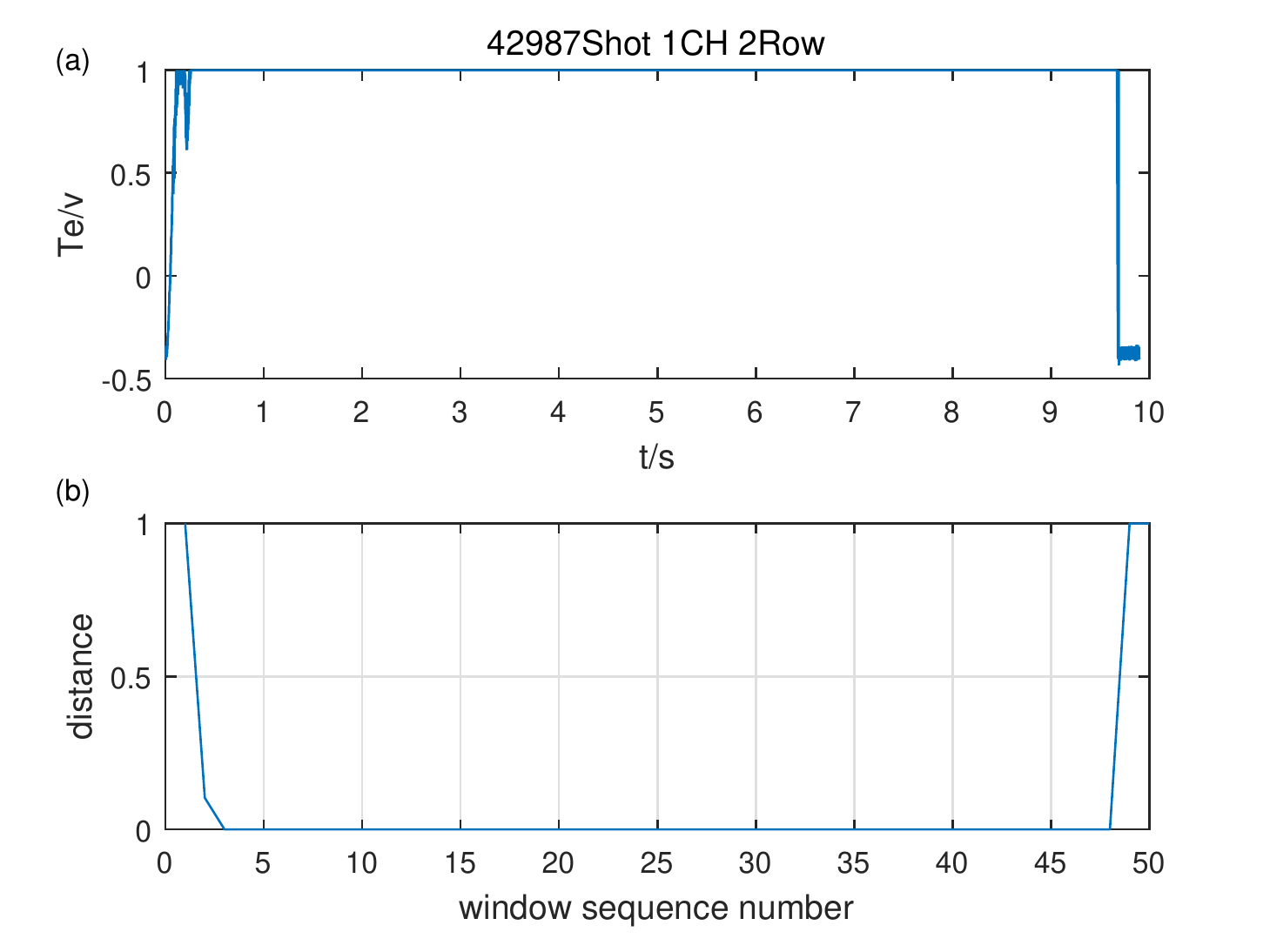}
\caption{\label{fig:3}(a) Raw signal of 42987th Shot 1CH 2Row (saturated).  (b) The signal of 42987th Shot 1CH 2Row is divided into fifty windows, and it shows the distance between each window and the range.}
\end{figure}

Figure~\ref{fig:3}(b) shows the distance of each window of 42987th Shot 1CH 2Row. After sorting, several minimum distances will be obtained. The feature-space of SVM consists of the minimum distance and two adjacent distances.

\subsubsection{Method of SVM for training data}
A random sampling of 20\% instances are taken as a test set, and the rest is used for training. It is found that when polynomial kernel functions are used, the model can be best trained and most of saturated signals are successfully identified. Only a few predictions are wrong, and they are found to be very similar to saturation signals.

\subsection{Identification of zero signals}
The feature of zero signals is that $T_{e}$ profiles of the steady segment are close to 0 V. After the identification of saturated signals, the remaining signals will be classified by the decision
tree algorithm. The distance between the steady segment and the noise fragment of the signal is selected as the feature-space of the decision tree algorithm. Signals are marked as zero signals and non-zero signals, respectively recorded as 2 and 0.

\subsubsection{Locating the steady segment}
Finding the steady segment of a signal is the precondition for the classification. In general, the steady segment is the smoothest segment within two seconds of the highest peak. The smoothness of the segment is represented by the standard deviation of $T_{e}$ profiles. The procedure of finding the steady segment is to locate the highest peak first and then compare the smoothness between segments around the highest peak. An example is given below.

In the same way, the signal is divided into fifty windows. And then the time average electron temperature for each window (< $T_{e}$ >) is obtained. The largest <$T_{e}$> is found and the corresponding window is marked as "C". Then three adjacent windows are grouped as one set, and there are 48 groups in all labeled by $S_1$, $S_2$, $S_3$......$S_{48}$.  Next, we calculate the smoothness of each group. Since the steady segment may appear on either side of the highest peak,  it is necessary to start at the C-th window, and compare the sum of the smoothness of left nine groups and right nine groups. The steady segment of the signal is in the smoother side. The assumption is that the right side is smoother. From $S_c$, $S_{c+1}$......$S_{c+9}$, the minimum value is found and identified the steady segment of this signal.

\begin{figure}[htbp]
\centering 
\includegraphics[width=.8\textwidth]{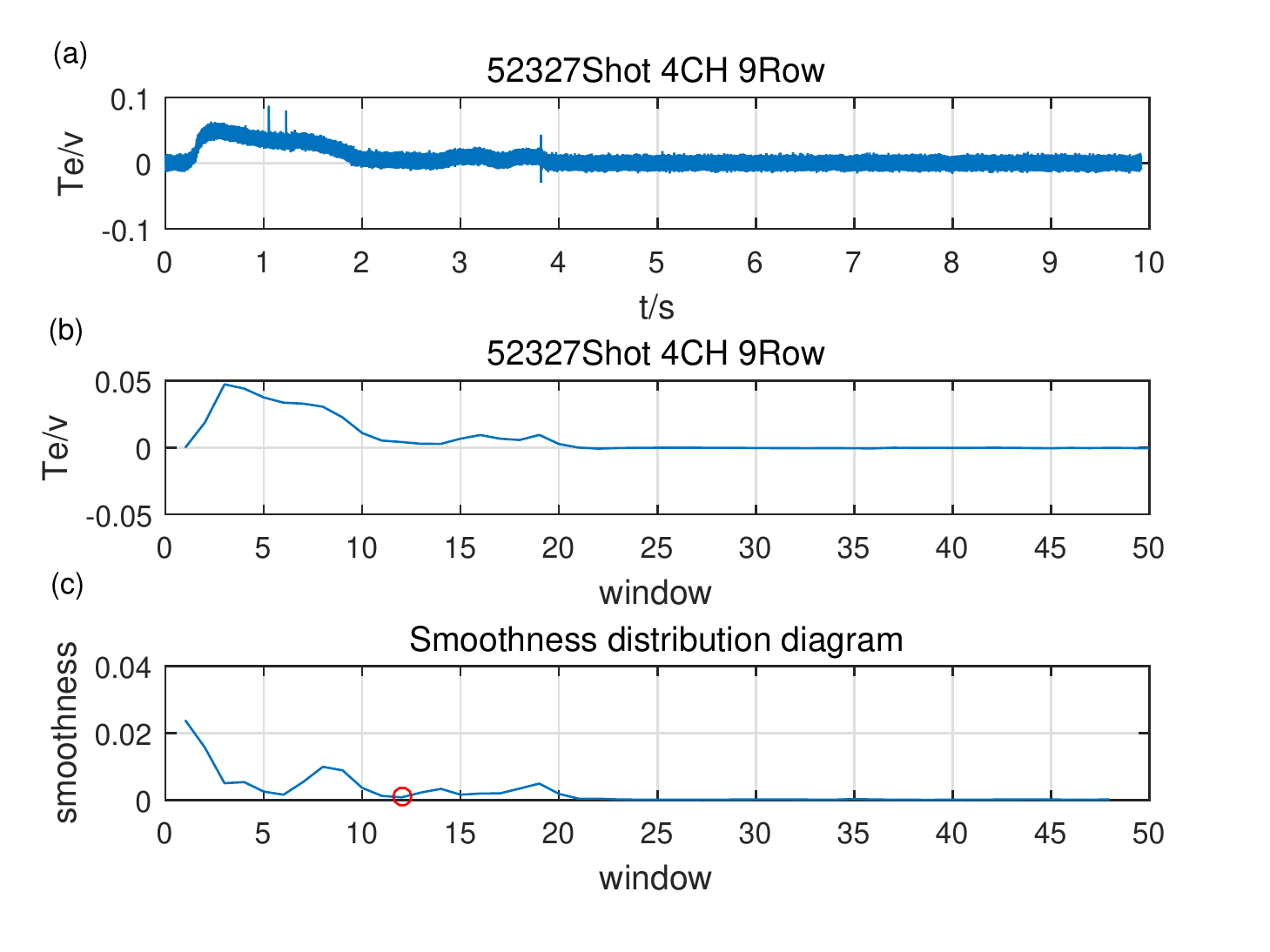}
\caption{\label{fig:4}(a) Raw signal of 52327 Shot 4CH 9Row which is a typical zero signal. (b) After
smoothing, the data of each window is averaged. (c) Smoothness distribution diagram is obtained by averaging over the fluctuation
of each window, and the 12-th window which is located at
the red marker is the steady segment.}
\end{figure}

Figure~\ref{fig:4}(b) shows  the <$Te$>  profiles of 52327 Shot 4CH 9Row. It is obvious that the highest peak lies in the third second. Figure~\ref{fig:4}(c) shows the smoothness of the entire signal. From $S_3$, $S_4$......$S_{12}$, the smallest can be quickly found. In Figure~\ref{fig:4}(b), the 12-th window which is located at the red marker in Figure~\ref{fig:4}(c) is indeed the steady segment.  There are 384 steady segments of one shot for each channel. In general, steady segments of most channels have the same position. The steady segment is composed of three adjacent windows. In order to reduce the error, the middle window is selected as the representative for the steady segment.

\subsubsection{Identify the noise at the end of the signal}
\label{sec:steady}
\begin{figure}[htbp]
\centering 
\includegraphics[width=.8\textwidth]{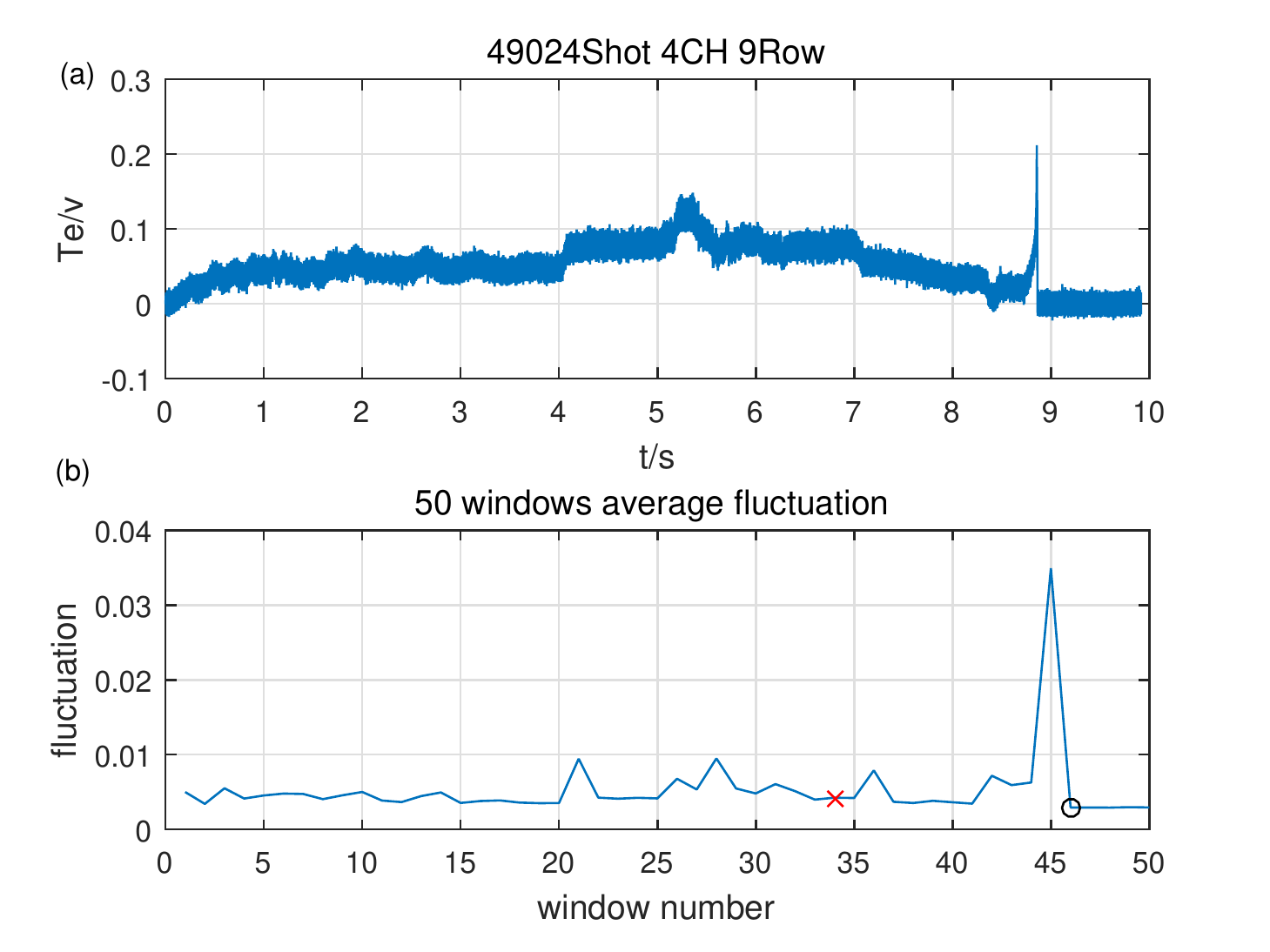}
\caption{\label{fig:5}(a) Raw signal 49024 Shot 4 CH 9Row. (b) $\sigma T_e$ profiles of 49024 Shot 4CH 9Row. The 34-th window which is marked by a red cross is the steady segment and the 46-th window which is labeled by a black circle is where the background noise begins.}
\end{figure}

A parameter of the model is the distance between the steady segment and the noise segments. So it is also important to find out noise segments. At the end of each signal, there will be a period of segment which contains almost noise. Figure~\ref{fig:5}(a) is the raw signal of 49024 Shot 4CH 9Row. Figure~\ref{fig:5}(b) shows $\sigma T_e$ profiles of fifty windows. The 34-th window which is marked by a red cross is the steady segment and the 46-th window which is labeled by a black circle is where the background noise begins. It can be seen from Figure~\ref{fig:5}(b) that a prominent peak sits between the steady segment and background noise segments. This is not a coincidence, but an inevitable stage for each signal. Before 9s there is active plasma emitting cyclotron radiation, and then the plasma terminates. Since it is a transient process, $\sigma T_e$ profiles will change very fast. This feature can be used to quickly identify the background noise of each signal. 

\subsubsection{Method of decision tree for training samples}
\begin{figure}[htbp]
\centering 
\includegraphics[width=.8\textwidth]{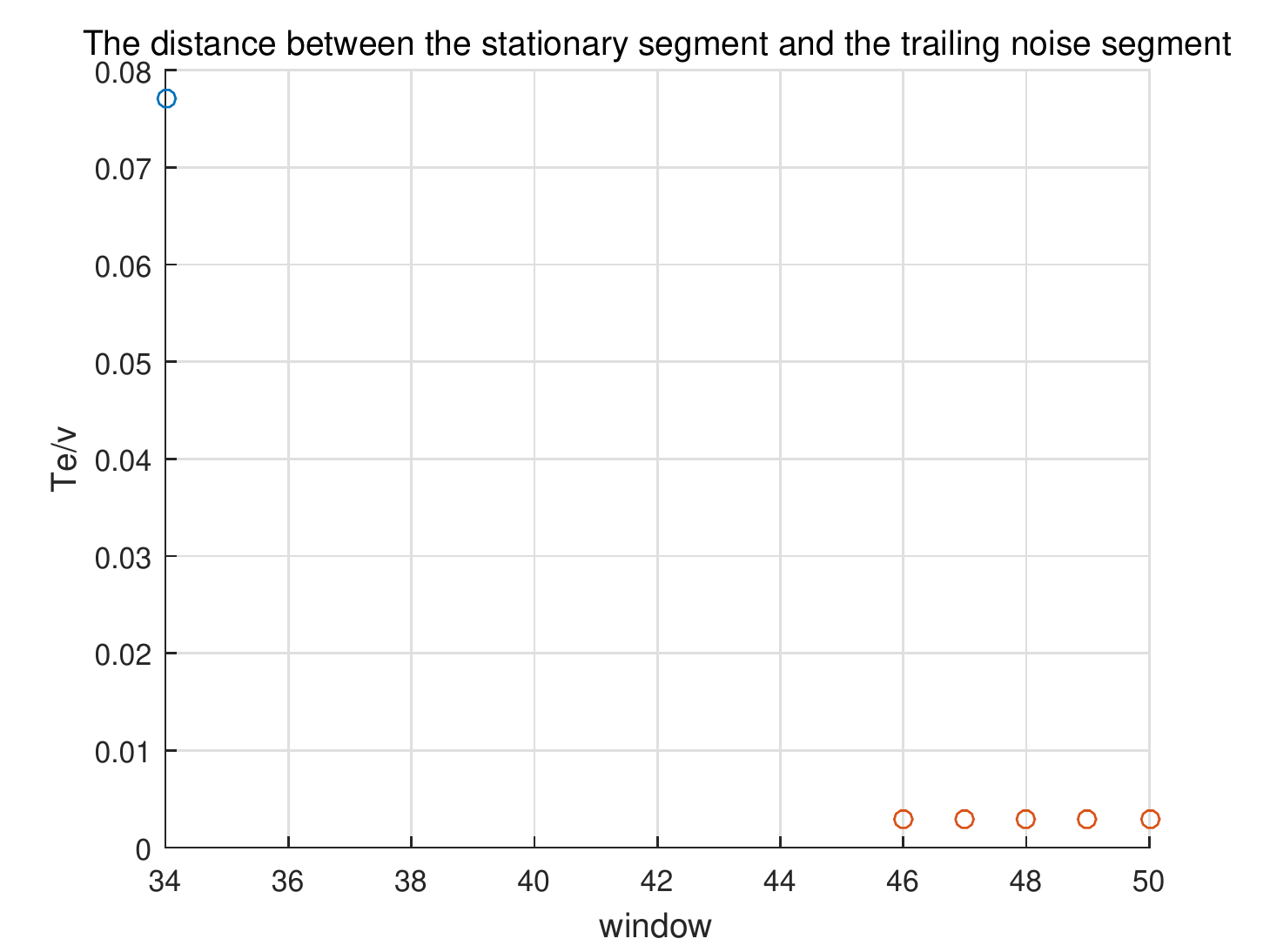}
\caption{\label{fig:6}The distance S between the steady segment and the trailing noise segment (49024 Shot 4CH 9Row). The blue marker represents the average of the smooth segment signals, i.e.,  < $T_{e1}$ >. The red markers represent the sum of the mean of the noise segment signal and the average of the fluctuation values, i.e., < $T_{e2}$ >+$\sigma T_{e2}$. If < $T_{e1}$ >-(< $T_{e2}$ >+$\sigma T_{e2}$) is small enough, the signal is identified as a zero signal.}
\end{figure}

According to the method described in Section~\ref{sec:steady}, the background noise of each signal is found. The next step is to calculate the distance S between the steady segment and background noise segments. 
In Figure~\ref{fig:6}, the blue marker is the time average temperature of the steady segment (< $T_{e1}$ >). The red marker indicates the sum of the time average temperature and the temperature fluctuation of the noise segments, i.e., < $T_{e2}$ > +$\sigma T_{e2}$. If < $T_{e1}$ > - (< $T_{e2}$ >+$\sigma T_{e2}$) is small enough, they will be classified as zero signals. Thus it is the parameter of decision tree to identify zero signals. In Figure~\ref{fig:6}, the distance S between the steady segment and noise segments of 49024 Shot 4CH 9Row is approximately 0.075 V, which is small. Thus, it is identified as a zero signal.

\subsection{Identification of weak signals}
The Signal-to-Noise Ratio (SNR) of weak signals resides between normal signals and zero signals. In the physical analysis of the ECEI data, the focus is on the relative temperature fluctuation of electrons ($\sigma T_e$/< $T_e$ >)~\cite{16}. In the actual measurement, $\sigma T_e$ profiles include both the normal electron temperature fluctuation and the background noise. If the SNR is too poor, the subsequent physical research will not make sense. Therefore it is necessary to identify weak signals. The noise of the system is generally inferred by comparing experiments with or without RF input~\cite{17}. It is preferable to use noise sections at the end of the signal as the background noise. The electron temperature fluctuation of the steady segment divided by that of noise segments ($\sigma T_{e1}$/$\sigma T_{e2}$ ) is selected as a parameter that represents the signal-to-noise ratio, and will be used as the parameter for the decision tree.

\section{Sample set}
\label{sec:4}
\begin{figure}[htbp]
\centering 
\includegraphics[width=.8\textwidth]{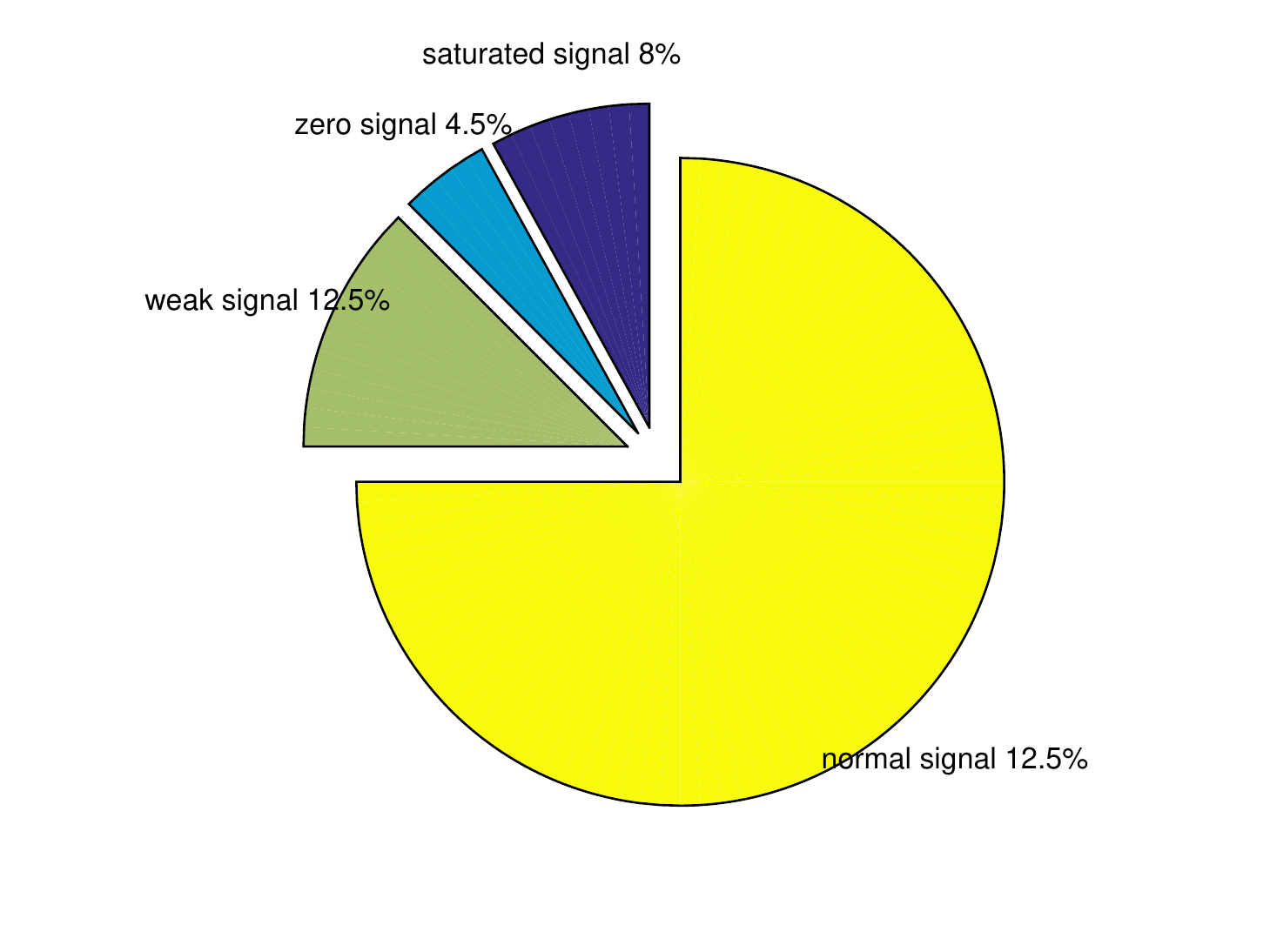}
\caption{\label{fig:7}The sample set for the classification.}
\end{figure}

Figure~\ref{fig:7} shows the proportion of saturated signals, normal signals, weak signals, and zero signals in the sample set. The 42987-th, the 42999-th, the 49024-th, and the 51064-th shot
are used as a sample set. They are tagged manually. Some of them are used for training, and the rest are used for testing. A total of 1536 samples are collected. 8\% of 1536 samples are saturated signals, 4.5\% are zero signals, 12.5\% are weak signals, and 75\% are normal signals. The sample set is randomly distributed. The proportion of abnormal signals is moderate and representative.  
First, after saturated signals are classified, 1409 signals are left. They include 61 zero signals and 1348 non-zero signals. After zero signals are identified, the rest are sent for the classification of weak signals and non-weak signals.

\section{Results of the automatic classification of experimental data}
\label{sec:5}
After the SVM and decision tree algorithms are implemented, massive raw data from the experiments of ECEI on EAST tokamak are fed into the learner for training, through which parameters of the model, such as window size and kernel function, have been optimized. The SVM and decision tree models with the optimized parameters generated satisfactory classification results. 

\subsection{The result of the identification of saturation signals}
In multiple tests, recognition rates of saturated signals reach 100\%. One of samples predicted wrong is the signal of 42987-th shot 22CH 3Row. It is found that this signal is very close to saturated signals. In order to test the effect of the model, a five-fold cross validation is adopted. 1536 samples are randomly divided into five groups. One of which is selected as the test sample and the other four are training samples. Sensitivity (Sen), specificity (Spe), and total accuracy (Q) are calculated respectively for each validation as follows,
\begin{equation}
\label{eq:2}
Sen = TP/(TP + FN),
\end{equation}
\begin{equation}
\label{eq:3}
Spe = TN/(TN + FP),
\end{equation}
\begin{equation}
\label{eq:4}
Q = (TP + TN)/(TP + FN + TN + FP).
\end{equation}

\renewcommand\arraystretch{1}
\begin{table}[htbp]
\centering
\caption{\label{tab:1} Results of five-fold cross validation for saturated signals.}
\smallskip
\begin{tabular}{cccccccc}
\hline
&TP&FN&TN&FP&Sen(\%)&Spe(\%)&Q(\%)\\
\hline
Cross-validation 1 & 26 & 0 & 280 & 1 & 100 & 99.64 & 99.67\\
Cross-validation 2 & 17 & 1 & 287 & 2 & 94.4 & 99.3 & 99\\
Cross-validation 3 & 24 & 1 & 282 & 1 & 96 & 99.65 & 99.35\\
Cross-validation 4 & 27 & 0 & 279 & 1 & 100 & 99.64 & 99.67\\
Cross-validation 5 & 26 & 0 & 279 & 2 & 100 & 99.29 & 99.35\\
total & 120 & 2 & 1407 & 7 & 98.36 & 99.5 & 99.4\\
\hline
\end{tabular}
\end{table}

The results are listed in  Table~\ref{tab:1}. Here, TP represents the number of saturated signals identified  correctly, FN represents the number of saturated signals that are wrongly identified as unsaturated signals, TN is number of unsaturated signals that are correctly classified, and FP is the number of unsaturated signals that are wrongly identified as saturated signals~\cite{18}. From table~\ref{tab:1}, Sen, Spe, and Q are almost 100\%. It is obvious that the model of the classification can accurately identify saturation signals.

\subsection{The result of the identification of zero signals}
1409 signals are randomly divided into five groups for cross-validation.  One of them is taken as a test sample, and the other four groups are learning samples. The results are listed in  Table~\ref{tab:2}, where TP is the number of zero signals correctly identified, FN represents the number of zero signals identified as nonzero signals, TN is the number of nonzero signals classified correctly, and FP stands for the number of nonzero signals classified as zero signals. From table~\ref{tab:2}, results of five-fold cross validation show few errors. The accuracy rate of identifying zero signals reaches 99.86\%.

\begin{table}[htbp]
\centering
\caption{\label{tab:2} Results of five-fold cross validation for zero signals.}
\smallskip
\begin{tabular}{cccccccc}
\hline
&TP&FN&TN&FP&Sen(\%)&Spe(\%)&Q(\%)\\
\hline
Cross-validation 1 & 17 & 0 & 264 & 0 & 100 & 100 & 100\\
Cross-validation 2 & 13 & 1 & 267 & 2 & 92.86 & 100 & 99.64\\
Cross-validation 3 & 10 & 0 & 271 & 0 & 100 & 100 & 100\\
Cross-validation 4 & 10 & 0 & 271 & 0 & 100 & 100 & 100\\
Cross-validation 5 & 9 & 1 & 275 & 0 & 90 & 100 & 99.65\\
total & 59 & 2 & 1348 & 0 & 96.7 & 100 & 99.86\\
\hline
\end{tabular}
\end{table}

\subsection{The result of the identification of weak signals}
\begin{table}[htbp]
\centering
\caption{\label{tab:3} Results of five-fold cross validation for weak signals}
\smallskip
\begin{tabular}{cccccccc}
\hline
&TP&FN&TN&FP&Sen(\%)&Spe(\%)&Q(\%)\\
\hline
Cross-validation 1 & 33 & 0 & 235 & 0 & 100 & 100 & 100\\
Cross-validation 2 & 27 & 0 & 241 & 0 & 100 & 100 & 100\\
Cross-validation 3 & 39 & 0 & 228 & 1 & 100 & 99.6 & 99.6\\
Cross-validation 4 & 30 & 0 & 238 & 0 & 100 & 100 & 100\\
Cross-validation 5 & 41 & 0 & 235 & 0 & 100 & 100 & 100\\
total & 170 & 0 & 1177 & 1 & 100 & 99.9 & 99.9\\
\hline
\end{tabular}
\end{table}

Based on results of statistics, the boundary between weak signals and non-weak signals is $\sigma T_{e1}$/$\sigma T_{e2}$ = 1.299. From table~\ref{tab:3}, results of five-fold cross validation are close to 100\%. One of samples predicted wrong is 49024-th shot 23CH 11Row whose $\sigma T_{e1}$/$\sigma T_{e2}$ profile is 1.3004, which is very close to the boundary between weak signals and normal signals.

\section{Summary}
\label{sec:6}
In summary, artificial intelligence technologies are applied to classifying the massive ECEI data on the EAST tokamak.  As a pretreating procedure, the data are separated into different segments. SVM algorithm is used to identify saturated signals and the method of decision tree is applied to classifying zero signals and weak signals. The models are trained using the massive ECEI data on the EAST tokamak based on optimized model parameters. Cross-validation studies showed that the model can identify saturated signals, zero signals and weak signals with accuracies of  99.4\%, 99.86\%, and 99.9\% respectively. It proves that this model can be used in practice. In the future study, similar automatic classification techniques will be developed and applied to identifying physical modes, such as plasma instabilities from valid ECEI signals.  

\acknowledgments
This work was supported partly by National key research and development program under Grant Nos. 2016YFA0400600, 2016YFA0400601 and 2016YFA0400602, Anhui Provincial Natural Science Foundation under Grant Nos. 1808085MA25 and also by the Fundamental Research Funds for the Central Universities with the Grant Nos. WK2150110008, Wk2030040098.


\begin{thebibliography}{99}
\bibitem{a}
Gao BX, Xie JL, Mao Z, Luo C, Zhu YL, Zhao ZL, Tong L, Liu WD, Luhmann NC, Domier CW and Tobias B, \emph{The electron cyclotron emission imaging system on EAST with continuous large observation area}, \emph{Journal of Instrumentation} {\bf 13(02)} (2018) P02009.

\bibitem{b}
Kim JB,  Lee W, Yun GS, Park HK, Domier CW and Luhmann NC Jr, \emph{Data acquisition and processing system of the electron cyclotron emission imaging system of the KSTAR tokamak}, \emph{Review of Scientific Instruments} {\bf 81(10)} (2010) 10D931.
\bibitem{c}
Baonian Wan for the EAST and HT-7 Teams and International Collaborators, \emph{Recent experiments in the EAST and HT-7 superconducting tokamaks}, \emph{Nuclear Fusion} {\bf 49(10)} (2009) 104011.
\bibitem{4}
M.Becoulet, M. Kim, G. Yun et al., \emph{Non-linear MHD modelling of edge localized modes dynamics in KSTAR}, \emph{Nuclear Fusion} {\bf 57(11)} (2017) 116059.
\bibitem{5}
BoWang, Bingjia Xiao, Jiangang Li, Yong Guo and Zhengping Luo, \emph{Artificial Neural Networks for Data Analysis of Magnetic Measurements on East}, \emph{Journal of Fusion Energy} {\bf 35} (2016) 390.
\bibitem{6}
N Isei, AIsayama, SIshida et al., \emph{Electron cyclotron emission measurements in JT-60U}, \emph{Fusion Engineering and Design} {\bf 53} (2001) 213-220.
\bibitem{7}
A. Vannucci, K.A. Oliveira and T. Tajima, \emph{Forecast of TEXT plasma disruptions using soft X rays as input signal in a neural network}, \emph{Nuclear Fusion} {\bf 39} (1999).
\bibitem{8}
T. Lan,J. Liu and H. Qin, \emph{Preference-based performance measures for Time-Domain Global Similarity method}, \emph{JINST} {\bf 12} (2017) C12008.
\bibitem{9}
Shi-ping Li, Fang-chao Chen and Long Wang, \emph{Modulation recognition algorithm of digital signal based on support vector machine}, \emph{Control and Decision Conference (CCDC)} (2012).
\bibitem{10}
Hongyan Zhao, \emph{The analysis and application of the C4.5 algorithm in decision tree technology}, \emph{Advanced Materials Research} {\bf 457-458} (2012) 754-757.
\bibitem{11}
Xu Xiao-Yuan, Wang Jun,  Yu Yi et al., \emph{Electron temperature fluctuation in the HT-7 tokamak plasma observed by electron cyclotron emission imaging}, \emph{Chinese Physics B} {\bf 18} (2009).
\bibitem{12}
Liu Yong, Ti Ang, Han Xiang et al., \emph{Present Status of the Electron Cyclotron Emission Measurements on HT-7 and EAST}, \emph{Plasma Science and Technology} {\bf 13(3)} (2011) 10090630.
\bibitem{13}
Zhao Z, Xie J, Qu C, Liao W, Li H, Lan T, Liu A, Zhuang G and Liu W, \emph{Analysis of sawtooth collapse time using electron cyclotron emission imaging on EAST tokamak}, \emph{Radiation Effects and Defects in Solids} {\bf 172(9-10)} (2017) 760-7.
\bibitem{14}
Azam Hussain, Zhenling Zhao, Jinlin Xie et al., \emph{Observations of compound sawteeth in ion cyclotron resonant heating plasma using ECE imaging on experimental advanced superconducting tokamak}, \emph{Physics of Plasmas} {\bf 23(4)} (2016) 042504.
\bibitem{15}
Azam Hussain, Gao Bing-Xi, Liu Wan-Dong and Xie Jin-Lin, \emph{Electron Cyclotron Emission Imaging Observations of m/n=1/1 and Higher Harmonic Modes during Sawtooth Oscillation in ICRF Heating Plasma on EAST}, \emph{Chinese Physics Letters} {\bf 32(6)} (2015) 065201.
\bibitem{16}
B.J.Tobias, R. L. Boivin, J. E. Boom et al., \emph{On the application of electron cyclotron emission imaging to the validation of theoretical models of magnetohydrodynamic activity}, \emph{Physics of Plasmas} {\bf 18} (2011) 056107.
\bibitem{17}
X. Han, X. Liu, Y. Liu et al., \emph{Design and characterization of a 32-channel heterodyne radiometer for electron cyclotron emission measurements on experimental advanced superconducting tokamak}, \emph{Review of Scientific Instruments} {\bf 85(7)} (2014) 10897623.
\bibitem{18}
Dinesh V. Rojatkar, Krushna D. Chinchkhede and G.G. Sarate, \emph{Handwritten Devnagari consonants recognition using MLPNN with five fold cross validation}, \emph{International Conference on Circuit, Power and Computing
Technologies} ICCPCT 2013.








\end{thebibliography}
\end{document}